\documentclass[conference]{IEEEtran}
\IEEEoverridecommandlockouts
\makeatletter
\def\@IEEEpubidpullup{5\baselineskip}  
\makeatother
\IEEEpubid{\begin{minipage}[b]{\textwidth}\ \\[12pt]
  \centering\small
  \textcopyright~2026 IEEE. Personal use of this material is permitted.
  Permission from IEEE must be obtained for all other uses, in any current
  or future media, including reprinting/republishing this material for
  advertising or promotional purposes, creating new collective works, for
  resale or redistribution to servers or lists, or reuse of any copyrighted
  component of this work in other works.
\end{minipage}}

\usepackage{cite}
\usepackage{amsmath,amssymb,amsfonts}
\usepackage{algorithmic}
\usepackage{graphicx}
\usepackage{textcomp}
\usepackage{xcolor}
\usepackage{booktabs}
\usepackage{url}
\usepackage{color, soul}
\def\BibTeX{{\rm B\kern-.05em{\sc i\kern-.025em b}\kern-.08em
    T\kern-.1667em\lower.7ex\hbox{E}\kern-.125emX}}
    
\begin{document}

\title{A Privacy-Preserving Framework Using Remote Data Science for Inter-Institutional Student Retention Prediction}

\author{%
\begin{tabular}[t]{@{}c@{\hspace{2em}}c@{\hspace{2em}}c@{}}
{\large John Fields} & {\large K M Sajjadul Islam} & {\large Ruchitha Thota} \\
\textit{Business Analytics} & \textit{Computer Science} & \textit{Computer Science}\\
\textit{Concordia University Wisconsin} & \textit{Marquette University} & \textit{Concordia University Wisconsin}\\
john.fields@cuw.edu & sajjad.islam@marquette.edu & ruchitha.thota@cuw.edu \\
\\
\multicolumn{3}{c}{%
\begin{tabular}{@{}c@{\hspace{5em}}c@{}}
{\large Victor Chen} & {\large Praveen Madiraju} \\
\textit{Computer Science} & \textit{Computer Science}\\
\textit{Georgetown University} & \textit{Marquette University}\\
vwc8@georgetown.edu & praveen.madiraju@marquette.edu
\end{tabular}}
\end{tabular}
}

\maketitle

\begin{abstract}
This study explores privacy-preserving machine learning (PPML) techniques using the PySyft platform to enable collaborative prediction of student retention between institutions. We developed a remote data science (RDS) framework with a semi-air-gapped architecture consisting of high-side and low-side servers, allowing researchers from three universities to build predictive models on sensitive student data without direct data access. Using historical data from a small private university (N=720), we evaluated three synthetic data generation approaches and validated the framework through inter-institutional collaboration. The results demonstrate consistent classification performance across institutions (Macro F1: 0.690--0.695) while maintaining strict Family Educational Rights and Privacy Act (FERPA) compliance. We also propose Data-Type-Aware Templates, a novel synthetic data method that prioritizes privacy over distributional fidelity. Our findings confirm that RDS-based PPML is technically feasible for educational settings and offers a practical alternative to federated learning for small-scale inter-institutional collaborations. The code is available at https://github.com/jtfields/NAIRR240195-Privacy-Preserving-Machine-Learning.
\end{abstract}

\begin{IEEEkeywords}
privacy-preserving machine learning, student retention, remote data science, PySyft, higher education
\end{IEEEkeywords}

\section{Introduction}

Student retention remains one of the most persistent challenges in higher education. Despite more than 70 years of research, completion rates have shown only modest improvements, with gains largely attributed to grade inflation and not to genuine educational advancement~\cite{denning2022}. This challenge affects both students, who face financial burdens as well as delayed career progression, and institutions that invest substantial resources in recruitment and support services.
\IEEEpubidadjcol 
Enhanced statistical methods and artificial intelligence have improved our ability to identify at-risk students~\cite{kemper2020, fields2024, alban2019, hinojosa2022}. These predictive approaches leverage academic performance, demographics, financial aid records, and campus engagement data to create early warning systems. However, a critical limitation pervades current research: most studies are confined to single institutions~\cite{gardner2019}. Models trained at one university may perform poorly at another, and smaller institutions often lack the data volume or technical expertise to develop robust models independently.

The fundamental barrier to multi-institutional collaboration lies in student data privacy. Educational records are protected by strict regulations such as FERPA in the United States and the General Data Protection Regulation (GDPR) in the European Union (EU), along with institutional policies that prevent data sharing between organizations. These restrictions, while essential for protecting student rights, create significant obstacles to collaborative research and inter-institutional model development.

This study extends our prior single-institution retention modeling work~\cite{fields2024} to a privacy-preserving, multi-institution setting and makes three contributions: (1) we demonstrate a PySyft-based remote data science framework that enables inter-institutional collaboration on student retention prediction while maintaining strict privacy compliance; (2) we propose Data-Type-Aware Templates, a novel approach to synthetic data generation that prioritizes privacy protection over distributional fidelity; and (3) we validate the framework through empirical evaluation with researchers from three universities of different sizes, providing insights into privacy-utility trade-offs.

\section{Related Work}

\subsection{Machine Learning (ML) for Student Dropout Prediction}

Educational data mining has demonstrated the effectiveness of machine learning in identifying at-risk students. Dekker et al.~\cite{dekker2009} found decision trees particularly effective for predicting first-year dropout. M\'{a}rquez-Vera et al.~\cite{marquez2016} advanced this with the Interpretable Classification Rule Mining algorithm for high school students. More recently, deep learning approaches have shown promise: Mubarak et al.~\cite{mubarak2021} developed Long Short Term Memory (LSTM) models for Massive Open Online Courses (MOOCs), while Albreiki et al.~\cite{albreiki2023} applied graph convolutional networks for at-risk student identification.

\subsection{Inter-Institutional Collaboration}

Most retention studies focus on individual institutions, limiting generalizability and hindering benchmarking~\cite{gardner2019}. Institutional contexts vary considerably in demographics, programs, and support systems~\cite{gasevic2016}, while limited sample sizes often do not capture population diversity.

Gardner et al.~\cite{gardner2023} studied inter-institutional model sharing across four U.S. universities (N = 1,000 to 30,000), comparing direct transfer, ensemble averaging, and hybrid approaches and finding that simple prediction averaging matched locally trained models. Our work extends this line of research to a smaller university (N=720) and uses remote data science in place of federated learning, exchanging data-owner-reviewed aggregate metrics or differentially privatized row-level outputs in place of raw per-record predictions. Table~\ref{tab:priorwork} summarizes the methodological differences with closely related prior work.

\begin{table}[t]
\centering
\caption{Comparison with Closely Related Prior Work}
\label{tab:priorwork}
\begin{tabular}{p{1.4cm}l p{1.5cm}p{1.8cm}}
\toprule
\textbf{Study} & \textbf{Scale (N)} & \textbf{Privacy Mechanism} & \textbf{Cross-site Exchange} \\
\midrule
Gardner et al.~\cite{gardner2023} & \shortstack[l]{4 univ.\\(N = 1k--30k)} & Federated, no raw data sharing & Per-record predictions / model outputs \\
\addlinespace
\textbf{This work} & \shortstack[l]{1 univ.\ (N = 720)\\+ 3 collab.\ univ.} & RDS + human-in-the-loop review & Aggregate metrics or DP-protected row-level outputs \\
\bottomrule
\end{tabular}
\end{table}

\subsection{Privacy-Preserving Machine Learning}

Two primary paradigms have emerged for PPML: Federated Learning (FL) and Remote Data Science (RDS). FL trains models across decentralized sites without centralizing data, sharing only model updates~\cite{ingerman2019}. However, FL assumes that all participants can support local training infrastructure and introduces risks of model inversion~\cite{fredrikson2015} and membership inference~\cite{shokri2016} attacks. Differential privacy~\cite{dwork2006} offers complementary mathematical guarantees, though real-world implementations often fall short~\cite{vanhaastrecht2024}. For educational research with modest numbers of universities, heterogeneous capabilities, and smaller sample sizes, FL's computational overhead may outweigh benefits.

RDS enables researchers to submit code to remote datasets, returning only non-disclosive aggregate results. This paradigm aligns naturally with educational research that involves 2--10 collaborating institutions. DataSHIELD~\cite{avraam2025} is the most established open-source RDS framework but operates within the R ecosystem, which limits integration with modern Python-based ML tools.

PySyft, developed by the OpenMined community, provides Python-native RDS with support for PyTorch, TensorFlow, scikit-learn, and pandas; integrated differential privacy mechanisms; domain-based access controls; and interactive Jupyter workflows~\cite{trask2020}. We selected PySyft for these capabilities, its flexibility across modeling approaches, the active OpenMined community, and native support for future deep learning architectures. Table~\ref{tab:comparison} summarizes the key differences between FL and RDS approaches.

\begin{table}[t]
\centering
\caption{Federated Learning vs. Remote Data Science}
\label{tab:comparison}
\begin{tabular}{p{1.6cm}p{2.8cm}p{2.8cm}}
\toprule
\textbf{Feature} & \textbf{Federated Learning} & \textbf{RDS (PySyft)} \\
\midrule
Data location & Remains at institution & Remains at institution \\
Movement & Model updates to aggregator & Code sent, results returned \\
Privacy risk & Moderate (model inversion possible) & Low (admin mediates output) \\
Typical scale & Many participants & 2--10 institutions \\
\bottomrule
\end{tabular}
\end{table}

\section{Methodology}

\subsection{Research Design}

This study evaluates the feasibility and effectiveness of PPML for inter-institutional student dropout prediction using RDS. We address three research questions: (1) Can synthetic data provide sufficient utility for model development before deployment on private data? (2) Does PySyft's RDS architecture enable effective collaboration while maintaining data privacy? (3) What are the practical challenges of implementing PPML in educational settings?

Our experimental design involved three universities of different sizes collaborating to predict student retention within a distributed architecture that maintained strict data isolation: Concordia University Wisconsin and Ann Arbor (approximately 6,500 students; the data-owning institution), Marquette University (approximately 12,000), and Georgetown University (approximately 20,000). Although the private data set originates from a single institution, the inter-institutional aspect refers to researchers from all three universities independently developing and evaluating models within the shared RDS framework. We used a two-phase approach: validation on publicly available synthetic data (Faketucky~\cite{faketucky2017}), followed by deployment on confidential institutional data.

\subsection{Privacy Architecture}

\begin{figure}[t]
\centering
\includegraphics[width=\columnwidth]{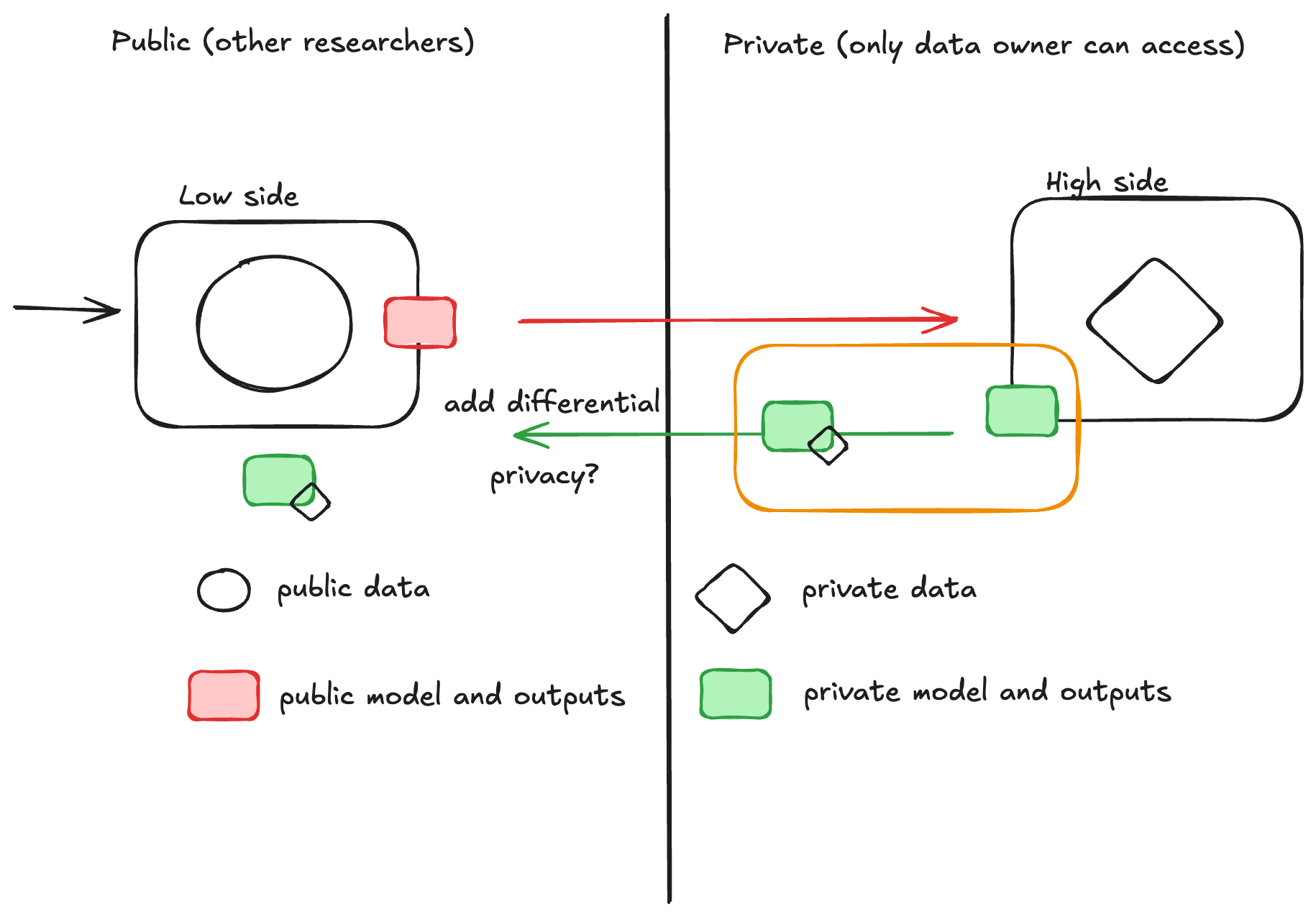}
\caption{Remote Data Science Architecture.}
\label{fig:arch}
\end{figure}

We implemented a two-server architecture consisting of a ``low-side'' public server and ``high-side'' secure server (Fig.~\ref{fig:arch}). The low-side server, hosted on Microsoft Azure through our National Artificial Intelligence Research Resource (NAIRR) grant partnership with OpenMined, contained synthetic data and served as the development environment accessible to all researchers. The high-side server, physically secured at Concordia University Wisconsin, housed confidential student records in a semi-air-gapped configuration with restricted network access via dedicated hardware.

This architecture enables the RDS workflow: data scientists develop and test models on the low-side using synthetic data, submit analysis code for data owner review, and receive outputs either as non-disclosive aggregate statistics or, when row-level outputs are required, with differential privacy applied.

Table~\ref{tab:specs} summarizes the server specifications. A dedicated administrator laptop on the university network synchronized approved code to the high-side server over SSH; the high-side server itself had no direct internet access.

\begin{table}[t]
\centering
\caption{Server Specifications}
\label{tab:specs}
\begin{tabular}{lll}
\toprule
\textbf{Component} & \textbf{High Side} & \textbf{Low Side} \\
\midrule
Hardware & Dell Precision T7610 & Microsoft Azure \\
OS & Ubuntu 24.04 & Ubuntu 24.04 \\
Python & 3.12.3 & 3.10.12 \\
PySyft & 0.9.2 & 0.9.2 \\
Docker & 24.0.9 & 28.0.4 \\
\bottomrule
\end{tabular}
\end{table}

\subsubsection*{FERPA Compliance Mapping}
FERPA restricts disclosure of personally identifiable information (PII) from education records absent written consent or a qualifying exception. Our framework satisfies these requirements as follows: (i) raw records never leave Concordia's high-side server, eliminating disclosure at the data layer; (ii) all returned outputs are non-disclosive: aggregate statistics directly or differentially privatized row-level results, reviewed by the data owner; (iii) PII fields are removed and remaining quasi-identifiers are de-identified prior to ingestion, consistent with FERPA \S 99.31(b) guidance; and (iv) the data owner exercises the school official function during code review, satisfying the legitimate-educational-interest provision. Institutional review board (IRB) approval and Chief Information Officer (CIO) oversight provide additional institutional accountability.

\subsection{Data and Prediction Task}

\subsubsection{Synthetic Data for Model Development}
The initial development of the model used the Faketucky synthetic education dataset~\cite{faketucky2017}, a publicly available resource designed to mimic the structures of real educational data. The prediction task involved binary classification of student retention (enrolled beyond the first year vs. departed), with the target variable showing class imbalance (18,776 retained, 66,384 departed, 26,831 missing). Records with missing target values were excluded from training and evaluation.

To evaluate synthetic data generation approaches, we compared three methods:

\textbf{Method 1 (SDV Gaussian Copula):} The Synthetic Data Vault (SDV) library~\cite{montanez2018} uses probabilistic graphical modeling and deep learning to generate artificial data preserving the statistical properties of the original. We applied SDV's Gaussian copula model to generate statistically similar synthetic records.

\textbf{Method 2 (SDV with Differential Privacy):} Identical to Method 1 but incorporating formal privacy guarantees through differential privacy ($\varepsilon = 1.0$, $\delta = 10^{-5}$) during synthesis.

\textbf{Method 3 (Data-Type-Aware Templates -- Proposed):} Instead of learning statistical distributions, this approach used the Python Faker library to generate structurally valid records based on variable types and ranges (e.g., grade point average (GPA) 0.0--4.0, American College Testing (ACT) composite scores 1--34, student age 18--25; valid category labels for gender, Pell eligibility, full-time status, college, degree program, ethnicity, and religion). Because the generated records have no statistical relationship to real data, this method provides the strongest privacy protection. For the RDS workflow, mock data need only be structurally valid to support model architecture development, since the final model is executed on real data by the data owner. Methods 1 and 2 were evaluated against Faketucky (Section~\ref{sec:results}); the production RDS deployment used only Method 3, since Method 1 leaks distributional information that could enable inference attacks and Method 2's differentially private outputs were unstable on this dataset.

\subsubsection{Private Institutional Data}
Following synthetic validation, we deployed the system using de-identified data from Concordia's 2021 cohort (N=720 students; 528 retained, 192 departed). Table~\ref{tab:datacomp} compares the structure of the retention variable between the Faketucky validation data and the institutional deployment data. The data set included 48 variables spanning demographics, academic preparation (ACT scores, high school grade point average (HS GPA)), enrollment patterns (credits attempted/earned, term GPA), financial aid indicators, and participation in institutional support programs.

\subsubsection{Model Configuration}
All models were implemented using scikit-learn classification algorithms, including logistic regression, random forest, and support vector machines. Data were split into training and testing sets using an 80--20 stratified split to preserve class distribution. Continuous features were standardized where applicable. Hyperparameters were selected using default settings with minor adjustments based on validation performance. All personally identifiable information was removed prior to analysis. The study received IRB approval with additional CIO oversight.

\begin{table}[t]
\centering
\caption{Retention Variable Comparison}
\label{tab:datacomp}
\begin{tabular}{lclc}
\toprule
\multicolumn{2}{c}{\textbf{Faketucky}} & \multicolumn{2}{c}{\textbf{Private Institution}} \\
\cmidrule(lr){1-2} \cmidrule(lr){3-4}
\textbf{Value} & \textbf{Freq.} & \textbf{Value} & \textbf{Freq.} \\
\midrule
Departed (0) & 66,384 & Departed (0) & 192 \\
Retained (1) & 18,776 & Retained (1) & 528 \\
Missing & 26,831 & Missing & 0 \\
\bottomrule
\end{tabular}
\end{table}

\subsection{Collaborative Workflow}

Three researchers from different universities independently developed classification models as shown in Fig.~\ref{fig:workflow}, proceeding through iterative cycles:

\begin{figure}[t]
\centering
\includegraphics[width=\columnwidth]{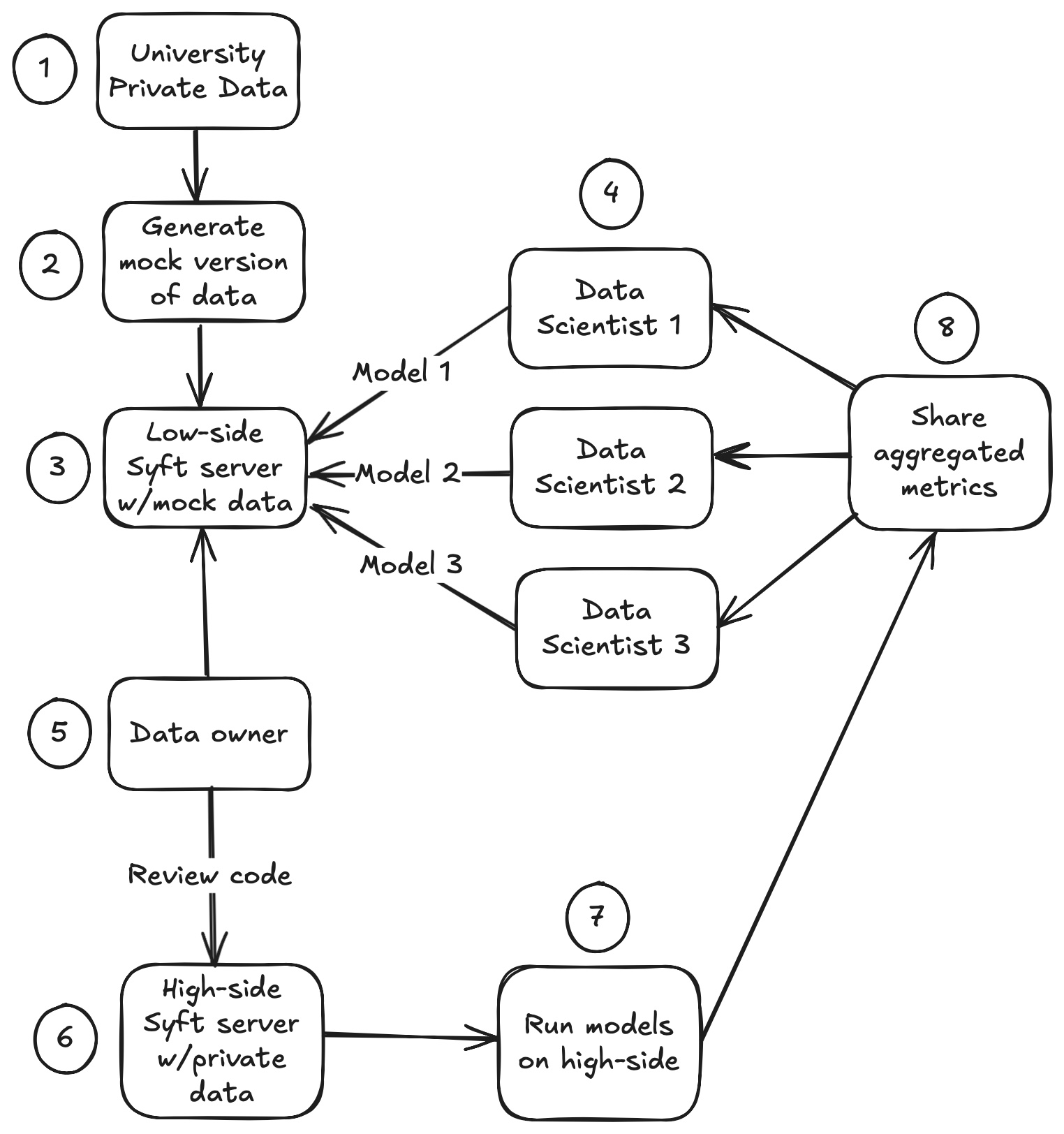}
\caption{Collaborative Workflow.}
\label{fig:workflow}
\end{figure}

\begin{enumerate}
\item The university's private data resides only on the high-side server.
\item The data owner generates mock data using Data-Type-Aware Templates (Method 3).
\item Mock data is deployed to the low-side Syft server for data scientists to explore and model.
\item Data Scientists 1--3 build models (Model 1--3) and test them on the mock data using scikit-learn, varying hyperparameters and feature engineering.
\item The data owner reviews each submission for (a) malicious code, (b) output-policy compliance (aggregate metrics returned directly; row-level outputs differentially privatized when required; raw data never released), and (c) technical validity.
\item Approved code synchronizes to the high-side Syft server.
\item Models execute on the high-side against the real private data.
\item Aggregate results (confusion matrices, F1 scores) return to the low-side; data scientists refine their models and repeat the cycle.
\end{enumerate}

This iterative process maintained strict data isolation; researchers never accessed raw data directly and outputs left the secure environment only after data-owner review (as aggregate statistics or differentially privatized row-level results).

\subsection{Threat Model and Privacy Assumptions}

We consider an adversary interacting with the system as an external researcher who submits analysis code and requests outputs. The data owner reviews each submission, returning aggregate statistics directly and applying differential privacy to row-level outputs when required; raw data are never released. The framework mitigates membership inference, attribute inference, and model inversion attacks through this human-in-the-loop enforcement combined with calibrated noise on row-level releases~\cite{shokri2016}. Table~\ref{tab:threat_model} summarizes the threat model.

\begin{table}[h]
\centering
\caption{Threat Model Summary}
\label{tab:threat_model}
\begin{tabular}{lp{5.5cm}}
\hline
\textbf{Component} & \textbf{Description} \\ \hline
Adversary & External researcher, honest-but-curious analyst \\
Access & Code submission; outputs returned as aggregate statistics or DP-protected row-level results \\
Restricted Access & Raw data never released; row-level outputs returned with differential privacy when required \\
Attacks Considered & Membership inference, attribute inference, model inversion \\
Defense Mechanism & Output restriction and human-in-the-loop review \\
\hline
\end{tabular}
\end{table}

\subsection{Privacy Challenges and Mitigations}

Several practical challenges emerged during deployment, each requiring targeted mitigations.

\textit{Row-level output requirements.} Georgetown's analysis required row-level classification weights, which by default would have created membership inference risk~\cite{shokri2016}. Instead of blocking the request, the data owner applied calibrated differential privacy noise to the row-level weights before release, illustrating that human-in-the-loop review enables flexible enforcement: aggregate statistics are returned directly, while legitimate row-level needs are accommodated through noise injection. We also enforced minimum cell-size thresholds during review so that low-frequency cells were suppressed before release.

\textit{Differential privacy instability.} Differential privacy mechanisms occasionally produced not-a-number (NaN) values when interacting with preprocessing steps, particularly when noise interacted with standardization on low-variance features. We addressed this through robust input validation (rejecting non-finite values before noise application), feature-wise epsilon budgeting (allocating separate privacy budgets per column instead of one global $\varepsilon$), and direct coordination with OpenMined engineering for upstream fixes.

\textit{Environment drift.} Package dependency mismatches between researcher environments and the Docker-based deployment caused intermittent execution failures. We mitigated this by providing documented setup instructions to participants, maintaining a shared base Docker image, and requiring a pre-submission validation step on the low-side mock data.

\subsection{Evaluation Metrics}

Model performance was assessed using precision, recall, F1-score, and accuracy. Beyond predictive accuracy, we evaluated synthetic-data utility (development-to-deployment performance transfer), operational feasibility (researcher feedback and incident documentation), and privacy protection (output review and disclosure risk analysis). All experimental code, configuration files, and documentation supporting reproduction of the reported experiments are available at https://github.com/jtfields/NAIRR240195-Privacy-Preserving-Machine-Learning.

\section{Results}
\label{sec:results}

\subsection{Synthetic Data Evaluation}

Models trained on SDV-generated synthetic data underperformed those trained on the original Faketucky data across every metric, with a mean degradation of 34.4\% (Table~\ref{tab:sdv}).

\begin{table}[t]
\centering
\caption{Classification Metrics: Original vs. SDV Synthetic Data (Faketucky)}
\label{tab:sdv}
\begin{tabular}{lccc}
\toprule
\textbf{Metric} & \textbf{With SDV} & \textbf{Without SDV} & \textbf{Decrease (\%)} \\
\midrule
Accuracy & 0.710 & 0.864 & 17.8 \\
F1-score & 0.487 & 0.824 & 40.9 \\
Precision & 0.559 & 0.786 & 28.8 \\
Recall & 0.432 & 0.866 & 50.1 \\
\midrule
Mean & 0.547 & 0.835 & 34.4 \\
\bottomrule
\end{tabular}
\end{table}

Standard SDV (Gaussian Copula) achieved the best fidelity, but models trained on synthetic data still underperformed on real test data. The synthetic data retain roughly two-thirds of the original model's effectiveness, with recall most affected (50.1\% drop), indicating that models become more conservative when trained on privacy-preserved data.

Statistical analysis confirms the degradation: a one-sample t-test yielded $t = 4.716$ ($df = 3$, $p \leq 0.05$), Cohen's $d = 2.187$ indicates a large effect, and variance analysis shows that privacy preservation significantly increases performance variability ($F = 15.42$, $p \leq 0.05$).

Adding differential privacy to SDV (Method 2, $\varepsilon = 1.0$, $\delta = 10^{-5}$) further degraded classification across all metrics, with recall showing the largest additional loss. Tightening the privacy budget below $\varepsilon = 1.0$ rendered models unusable on this dataset, consistent with reported privacy--utility frontiers for tabular educational data~\cite{vanhaastrecht2024}. Faker-generated mock data (Method 3) provided the strongest privacy protection of the three approaches, since it was entirely synthetic with no statistical relationship to real records. Its value lies in supporting the secure PPML workflow: external researchers prototype on structurally valid mock data, and the data owner then runs the finalized model on real records.

\subsection{Inter-Institutional PPML Results}

Table~\ref{tab:results} presents classification performance on private institutional data across the three participating universities, where class 0 represents non-returning students and class 1 represents returning students. These results were obtained through the Method 3 (Data-Type-Aware Templates) production workflow: the three researchers developed and validated their pipelines on structurally valid mock data on the low-side, after which the data owner executed the finalized models on the real private data on the high-side.

\begin{table}[t]
\centering
\caption{Researcher Classification Performance on Private Data}
\label{tab:results}
\begin{tabular}{lcccccc}
\toprule
& \multicolumn{3}{c}{\textbf{Class 0 (Departed)}} & \multicolumn{3}{c}{\textbf{Class 1 (Retained)}} \\
\cmidrule(lr){2-4} \cmidrule(lr){5-7}
\textbf{Institution} & \textbf{Prec} & \textbf{Rec} & \textbf{F1} & \textbf{Prec} & \textbf{Rec} & \textbf{F1} \\
\midrule
Concordia & 0.79 & 0.41 & 0.54 & 0.78 & 0.95 & 0.85 \\
Marquette & 0.70 & 0.42 & 0.52 & 0.82 & 0.93 & 0.87 \\
Georgetown & 0.85 & 0.37 & 0.52 & 0.77 & 0.97 & 0.86 \\
\bottomrule
\end{tabular}
\end{table}

The inter-institutional analysis reveals consistent overall performance, with Macro F1 scores ranging from 0.690 to 0.695 (coefficient of variation (CV) = 0.4\%). This aggregate consistency masks class-specific variation: Class 0 metrics show higher institutional variability (CV = 2.3--9.6\%) than Class 1 (CV = 1.2--3.2\%), reflecting the difficulty of predicting departure. Georgetown is most conservative, with the highest Class 0 precision (0.85) but lowest recall (0.37); all institutions achieve high Class 1 recall (0.93--0.97), reflecting the class imbalance (528 retained vs. 192 departed).

The successful execution of this framework demonstrates both technical feasibility and practical viability: researchers from three universities developed competitive models on private educational data without ever directly accessing it. Because the data owner executes the same scikit-learn pipelines on the high-side, the macro F1 reported here matches a centralized training run on the same data; the RDS cost is operational (review and submission cycles), not statistical. RDS and FL address model-sharing risks such as gradient leakage~\cite{fredrikson2015} through different mechanisms: FL typically applies differential privacy to gradients, while RDS uses human-in-the-loop review with differential privacy applied selectively to row-level outputs.

\section{Discussion}

\subsection{Key Findings and Contributions}

This study demonstrates that collaborative ML across institutions is technically feasible while maintaining FERPA compliance. The semi-air-gapped architecture with high-side and low-side servers proved effective for creating a secure collaborative environment. The successful participation of researchers from three distinct institutions validates the practical applicability of RDS for multi-institution collaboration.

Our synthetic-data evaluation revealed clear privacy-utility trade-offs. Standard SDV achieved reasonable fidelity, but models still underperformed on real data (mean degradation of 34.4\%). Our proposed Data-Type-Aware Templates method provides the strongest privacy guarantees while supporting the RDS workflow where external researchers prototype on structurally valid mock data before final models execute on private data, underscoring the value of RDS approaches that enable collaboration without requiring high-fidelity synthetic data.

The consistently low Class 0 recall (0.37--0.42) across all institutions reflects class imbalance in the dataset. Our prior work on a comparable cohort showed that SMOTE during training raised non-retained-class F1 scores from 0.00--0.13 to 0.82--0.89 across Gradient Boosting and XGBoost~\cite{fields2024}, suggesting substantial Class 0 gains are achievable when SMOTE and class-weighted ensemble methods are integrated into the RDS workflow. A systematic evaluation is left to follow-up work.

The study was intentionally constrained to scikit-learn classifiers, excluding neural network frameworks because they introduce additional privacy vulnerabilities through gradients and intermediate representations~\cite{fredrikson2015, shokri2016}; future implementations would require enhanced security protocols and more rigorous code review.

\subsection{Implications and Limitations}
The framework enables smaller institutions to benefit from collaborative model development without sharing sensitive records, and the modest hardware and software investment improves accessibility for institutions with limited IT resources. Beyond higher education, healthcare, government, and other regulated sectors could adapt this framework while maintaining regulatory compliance.

Practical barriers remain: server configuration complexity, Docker deployment instability, package dependency mismatches, and the need for cross-functional expertise. Emerging tools such as OpenMined's SyftBox~\cite{openmined2025} may reduce these barriers. PPML systems also remain vulnerable to multi-step disclosure risks where adversaries combine query results to infer individual information~\cite{fredrikson2015,shokri2016,dwork2006}, requiring layered defenses including differential privacy, minimum cell size requirements, and comprehensive logging.

\section{Conclusion}

This study demonstrates that PPML techniques can enable collaborative prediction of student retention across institutions while maintaining strict data privacy and regulatory compliance. The PySyft-based RDS framework was validated with researchers from three universities of different sizes, and current synthetic data methods remain insufficient for complex predictive tasks (mean degradation of 34.4\%), underscoring the importance of RDS approaches that enable collaboration without data sharing. Future work will incorporate class imbalance handling, explore neural network architectures with enhanced privacy safeguards, and extend the framework to multiple institutions contributing private data simultaneously.

\section*{Acknowledgment}

This work was supported through the NAIRR grant NAIRR 240195, Privacy-Preserving Machine Learning for Improving University Student Retention.

\section*{Declaration of Generative AI Tools}

The authors used ConnectedPapers.com, Consensus.app, and Claude.ai to brainstorm the Related Works section, Writefull for grammar checking, and ChatGPT to assist with code comments and boilerplate in some Jupyter notebooks. The authors reviewed, tested, and verified all generated content and take full responsibility for the publication.


\begin{thebibliography}{00}

\bibitem{denning2022} J.~T. Denning, E.~R. Eide, K.~J. Mumford, R.~W. Patterson, and M.~Warnick, ``Why have college completion rates increased?'' \textit{Am. Econ. J. Appl. Econ.}, vol.~14, no.~3, pp.~1--29, 2022.

\bibitem{kemper2020} L.~Kemper, G.~Vorhoff, and B.~U. Wigger, ``Predicting student dropout: A machine learning approach,'' \textit{Eur. J. Higher Educ.}, vol.~10, no.~1, pp.~28--47, 2020.

\bibitem{fields2024} J.~Fields, K.~Chovanec, and P.~Madiraju, ``Integrating categorical and continuous data in a cluster-then-classify methodology for predicting undergraduate student success,'' in \textit{Proc. IEEE Big Data Conf.}, 2024, pp.~8090--8098.

\bibitem{alban2019} M. Alban and D. Mauricio, ``Predicting university dropout through data mining: A systematic literature,'' \emph{Indian J. Sci. Technol.}, vol. 12, no. 4, pp. 1--12, 2019.

\bibitem{hinojosa2022} M.~Hinojosa \textit{et al.}, ``Student clustering procedure according to dropout risk to improve student management in higher education,'' \textit{Texto Libre}, vol.~15, 2022.

\bibitem{gardner2019} J.~Gardner, Y.~Yang, R.~Baker, and C.~Brooks, ``Modeling and experimental design for MOOC dropout prediction: A replication perspective,'' in \textit{Proc. EDM}, 2019.

\bibitem{dekker2009} G.~Dekker, M.~Pechenizkiy, and J.~Vleeshouwers, ``Predicting students drop out: A case study,'' in \textit{Proc. EDM}, 2009, pp.~41--50.

\bibitem{marquez2016} C.~M\'{a}rquez-Vera \textit{et al.}, ``Early dropout prediction using data mining: A case study with high school students,'' \textit{Expert Syst.}, vol.~33, no.~1, pp.~107--124, 2016.

\bibitem{mubarak2021} A.~A. Mubarak, H.~Cao, and S.~A.~M. Ahmed, ``Predictive learning analytics using deep learning model in MOOCs courses videos,'' \textit{Educ. Inf. Technol.}, vol.~26, no.~1, pp.~371--392, 2021.

\bibitem{albreiki2023} B.~Albreiki, T.~Habuza, and N.~Zaki, ``Extracting topological features to identify at-risk students using ML and GCN models,'' \textit{Int. J. Educ. Technol. Higher Educ.}, vol.~20, no.~1, 2023.

\bibitem{gasevic2016} D.~Ga\v{s}evi\'{c}, S.~Dawson, T.~Rogers, and D.~Gasevic, ``Learning analytics should not promote one size fits all,'' \textit{Internet High. Educ.}, vol.~28, pp.~68--84, 2016.


\bibitem{gardner2023}
J. Gardner, R. Yu, Q. Nguyen, C. Brooks, and R. Kizilcec, ``Cross-institutional transfer learning for educational models: Implications for model performance, fairness, and equity,'' in \emph{Proc. ACM Conf. Fairness, Accountability, and Transparency (FAccT)}, 2023, pp. 1664--1684.

\bibitem{ingerman2019} A.~Ingerman and K.~Ostrowski, ``Introducing TensorFlow Federated,'' TensorFlow Blog, 2019.

\bibitem{fredrikson2015} M.~Fredrikson, S.~Jha, and T.~Ristenpart, ``Model inversion attacks that exploit confidence information and basic countermeasures,'' in \textit{Proc. ACM CCS}, 2015.

\bibitem{shokri2016} R.~Shokri, M.~Stronati, C.~Song, and V.~Shmatikov, ``Membership inference attacks against machine learning models,'' in \textit{Proc. IEEE Symposium on Security and Privacy (S\&P)}, 2017, pp.~3--18.

\bibitem{avraam2025} D.~Avraam \textit{et al.}, ``DataSHIELD: Mitigating disclosure risk in a multi-site federated analysis platform,'' \textit{Bioinform. Adv.}, vol.~5, no.~1, 2025.

\bibitem{trask2020} A. Trask et al., ``Beyond privacy trade-offs with structured transparency,'' arXiv preprint arXiv:2012.08347, 2020.

\bibitem{montanez2018} A.~Montanez, ``SDV: An open source library for synthetic data generation,'' M.Eng. thesis, MIT, 2018.

\bibitem{faketucky2017} Center for Education Policy Research at Harvard University, ``Faketucky: OpenSDP college-going dataset,'' 2017.

\bibitem{openmined2025} OpenMined, ``SyftBox,'' 2025. [Online]. Available: \url{https://syftbox-documentation.openmined.org/}

\bibitem{vanhaastrecht2024} M.~van Haastrecht, M.~Brinkhuis, and M.~Spruit, ``Federated learning analytics: Investigating the privacy-performance trade-off,'' in \textit{LNCS}, Springer, 2024, pp.~62--74.

\bibitem{dwork2006} C.~Dwork, ``Differential privacy,'' in \textit{ICALP}, LNCS, Springer, 2006, pp.~1--12.

\end{thebibliography}
\end{document}